\documentclass[epj]{webofc}
\usepackage[varg]{txfonts}   
\usepackage{graphicx}
\usepackage{epsfig}
\usepackage{subfigure}
\usepackage{amsmath}
\wocname{EPJ Web of Conferences}
\woctitle{New Frontiers in Physics 2014}

\begin{document}
\selectlanguage{english}
\title{Search for exotics in the rare decay $B \rightarrow J/\psi K K K$ at $BABAR$.}

\author{Elisabetta Prencipe\inst{1,2}\fnsep\thanks{\email{e.prencipe@fz-juelich.de}}
        on behalf of the $BABAR$ Collaboration.
       }

\institute{Forschungszentrum J\"ulich, Leo Brandt strasse, 52428 J\"ulich, Germany 
\and
           \small{Previously addressed at the Johannes Gutenberg, University of Mainz, Germany} 
          }

\abstract{%
  One of the most intriguing puzzles in hadron spectroscopy are the
numerous charmonium-like states observed in the last decade, including charged states that are manifestly exotic. Over the years, the experiment $BABAR$ has extensively studied those in B meson decays, initial state radiation processes and two photon reactions. We report in this paper a new study  on some of those states, performed using the entire data sample collected by $BABAR$ in $e^+e^-$ collisions,  at center of mass energies near 10.58 GeV/c$^2$. The study of the process $B \rightarrow J/\psi \phi K$ will be presented, and the search for the resonant states X(4140) and X(4270) in their decays to  $J/\psi \phi$, will be highlighted. 
}
\maketitle

\section{Motivation}

$Strangeness$ in charmonium seems a sector still to be exploited.
While resonant structures like  the X(3872) have been seen in $B \rightarrow X K, X \rightarrow J/\psi~ \pi^+ \pi^-$, or like Y(4260) by investigating the process $e^+e^- \rightarrow \gamma_{ISR}X$, $X \rightarrow J/\psi \pi^+ \pi^-$\cite{X3872, X3872choi, Y4260babar}, no indication of new states has been observed in the  $J/\psi~ K^+ K^-$ invariant mass system, until the paper quoted in Ref.~\cite{kai} highlighted the possibility of a couple of resonant states, decaying to $J/\psi\phi$, with $\phi \rightarrow K^+ K^-$ and $J/\psi \rightarrow \mu^+ \mu^-$. These observations are nowaday controversial.

 The rare decay $B \rightarrow J/ \psi \phi K$ is interesting because it is a promising place to search for new resonances, as it proceeds, at quark level, via the weak transition $b \rightarrow c \bar c s$. It could be a quasi 2-body decay, $B \rightarrow X_g K$, with $X_g \rightarrow J/ \psi \phi$, where $X_g$ = $\lvert g c \bar c s \bar s>$, with gluonic contribution ($g$).

The QCD spectrum is much richer than that of the naive quark model, as the gluons, which mediate the strong force between quarks, can also act as principal components of entirely new types of hadrons. These gluonic hadrons fall into two general categories: glueballs (excited states of pure glue) and hybrids (resonances consisting largely of a quark, an antiquark, and excited glue). \\
We look for possible resonant states decaying to two mesons: $J/ \psi$, and  another meson  with strange $s$-quark content. The present analysis describes the case of $X_g \rightarrow J/\psi \phi$, $\phi \rightarrow K^{+}K^{-}$. We look also for exotic charged states, such as $Z \rightarrow J/\psi K^+$. $J/\psi$ and $\phi$ are 2 vector states, so non-parametrizable polarization effects can be shown in the dynamics of their interaction. Thus, more complications can arise compared to the phase-space (PHSP) model. Exotic quantum number combinations are theoretically allowed in this case. Predictions for hybrids come mainly from calculations based on the bag model, flux tube model, constituent gluon model and recently, with increasing precision, from Lattice QCD. 

We present a  new determination of the branching fraction (BF) of $B^{\pm,0} \rightarrow J/\psi K^+ K^- K^{\pm,0}$ and $B^{\pm,0} \rightarrow J/\psi \phi K^{\pm,0}$, using eight times more data than the previous analyses\cite{BBarArticle, cleoArticle}. A study of the $J/\psi \phi$ invariant mass spectrum is later reported, together with the invariant mass study of the $J/\psi K$ and the $K K K$ invariant mass systems. The upper limit (UL) of the decay $B^{0} \rightarrow J/\psi \phi$ is also presented.   

\section{Analysis strategy and results}

The analysis $B^+ \rightarrow J/\psi K^+K^- K^+$,  $B^0 \rightarrow J/\psi K^+K^- K^0_S$, $B^+ \rightarrow J/\psi \phi  K^+$, $B^0 \rightarrow J/\psi \phi  K^0_S$ and $B^0 \rightarrow J/\psi \phi$ are performed using 469 million $B \bar B$ pairs collected by $BABAR$. With $B^+$ we will imply in the text also the charged conjugate $B^-$.
 $B^+$ candidates are reconstructed by combining the $J/\psi$ candidate, reconstructed to $e^+e^-$ and $\mu^+ \mu^-$, with three loosely identified kaons. $J/\psi$ is  mass constraint. Similarly, for $B^0 \rightarrow J/\psi K^- K^+ K^0_S$ candidates, we combine the $J/\psi$ and $K^0_S$ with two loosely identified kaons and a $K^0_S$, which is formed by geometrically constraining a pair of oppositely charged tracks to a common vertex. $K^0_S$ is reconstructed to $\pi^+ \pi^-$. The $\phi(1020)$ signal region is selected  in the mass range [1.004; 1.034] GeV/c$^{\rm 2}$.  
%In Table~\ref{Table2-babar} the BF of the decays $B^+ \rightarrow J/\psi \phi K^+   $ and $B^0 \rightarrow J/\psi \phi K^0_S $ are reported: 
An unbinned  maximum likelihood fit is performed  to extract the yield and calculate the BFs. Detailed explanation on these calculations are presented in Ref.~\cite{elisabetta}, together with the relevant discussion for the non-resonant $K^+K^-$ contribution to the BF of $B \rightarrow J/\psi KKK$ and systematic uncertainty calculation. Here we report only the relevant information for the analysis of the three invariant mass distributions: $J/\psi \phi$, $J/\psi K$, $KKK$, for both charged and neutral B samples.

In this analysis we calculate also: 
$R_\phi$ = $\cal B$($B^0 \rightarrow J/\psi \phi K^0_S$)/$\cal B$($B^+ \rightarrow J/\psi \phi K^+$) = 0.48 $\pm$ 0.09 $\pm$ 0.02 , and $R_{2K}$ = $\cal B$($B^0 \rightarrow J/\psi K^+ K^- K^0_S$)/$\cal B$($B^+ \rightarrow J/\psi K^+ K^- K^+$) = 0.52 $\pm$ 0.09 $\pm$ 0.03; we find values in agreement with the expectation of the spectator quark model (e.g., ratio R$\sim$0.5). These are first measurements. For the first time the non-resonant $K^+K^-$ contribution to the BF of $B \rightarrow J/\psi K K  K$ is observed. \\
No evidence of signal is found for $B^0 \rightarrow J/\psi \phi$, in agreement with theoretical predictions: we evaluate UL<1.01 $\cdot$ 10$^{-6}$ at 90$\%$ confidence level (CL).

We search for the resonant states reported by the CDF Collaboration in the
$J/\psi \phi$ mass spectrum. The masses and the widths in our fit are fixed to values according to Ref.~\cite{kai}. 
\begin{figure}[htb]
\centering
\mbox{
{\scalebox{0.23}{\includegraphics{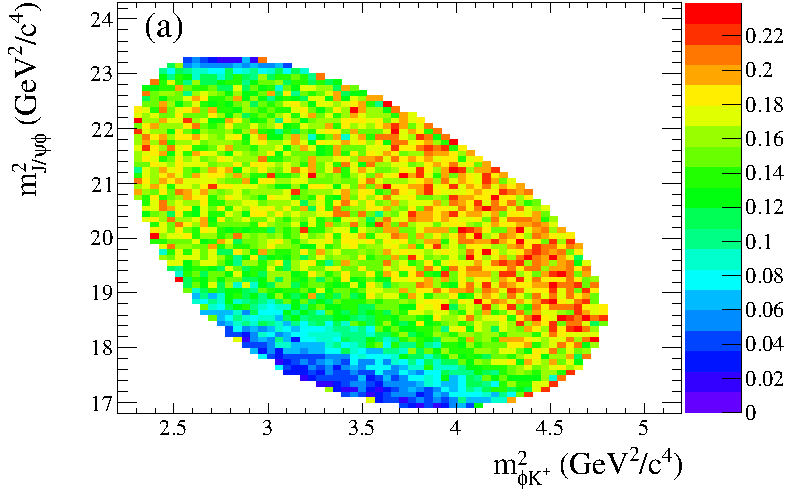}}} \quad
{\scalebox{0.23}{\includegraphics{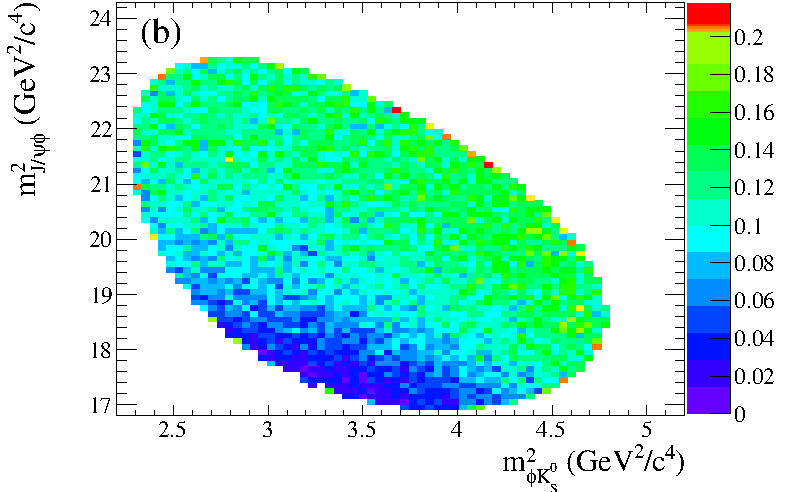}}} \quad
{\scalebox{0.185}{\includegraphics{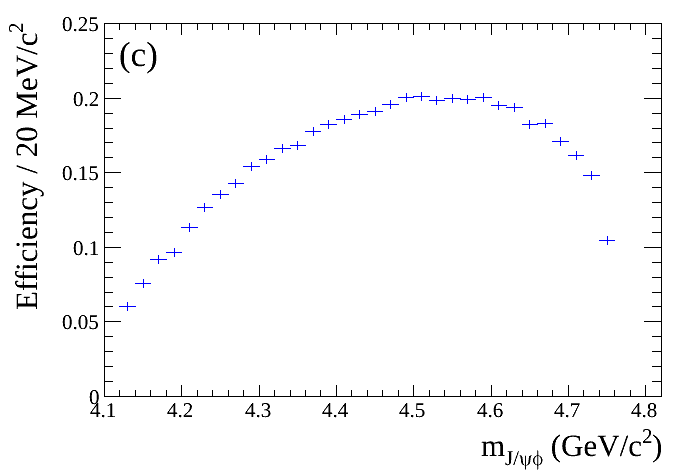}}}
}
\label{dalitz-babar}
\caption{ Efficiency distribution on the Dalitz plot for (a) $B^+ \rightarrow J/\psi \phi K^+$ and (b) $B^0 \rightarrow J/\psi \phi K^0_S$. (c) Average efficiency distribution as a function of the $J/\psi \phi$ mass for $B^+ \rightarrow J/\psi \phi K^+$.}
\end{figure}

\begin{figure}[htb]
\centering
\mbox{
{\scalebox{0.23}{\includegraphics{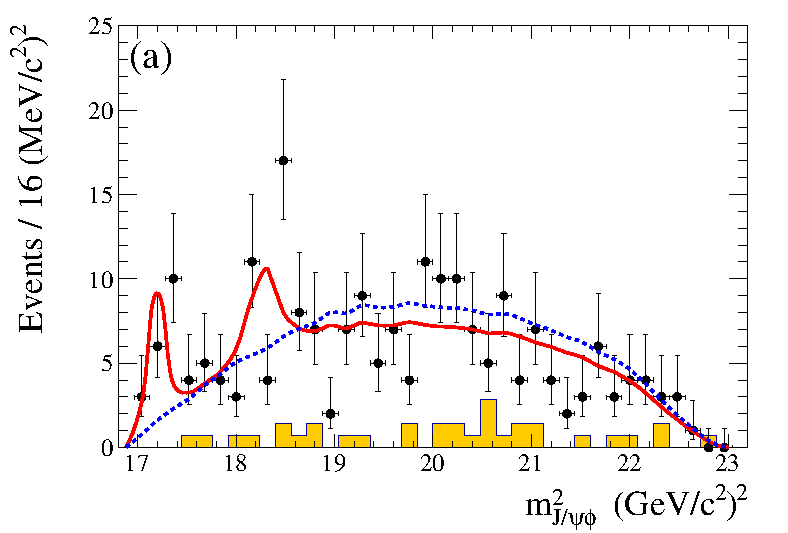}}} \quad
{\scalebox{0.19}{\includegraphics{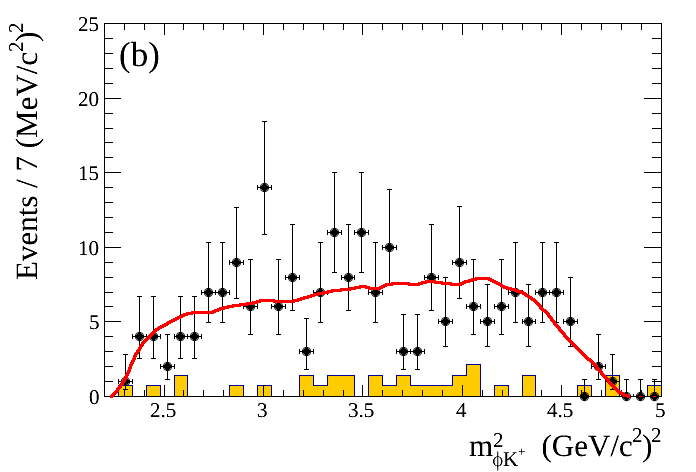}}} \quad
{\scalebox{0.23}{\includegraphics{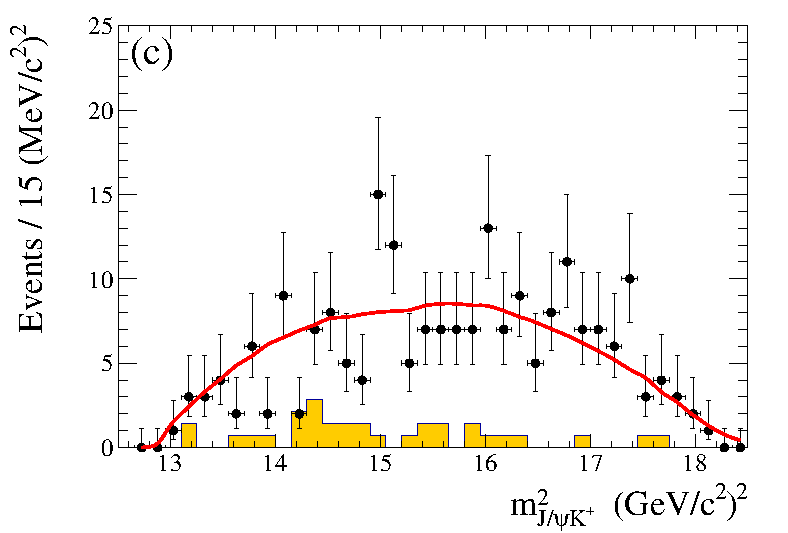}}}
}
\label{Fig20-babar}
\caption{Dalitz plot projections for $B^+ \rightarrow J/\psi \phi K^+$ on (a) $m^2_{J/\psi \phi}$, (b) $m^2_{\phi K^+}$, and (c) $m^2_{J/\psi K^+}$. The continuous (red) curves are the results from fit model performed including the $X(4140)$ and $X(4270)$ resonances. The dashed (blue) curve in (a)  indicates the projection for fit model  with no resonances included in the fit. The shaded (yellow) histograms indicate the evaluated background.}
\end{figure}
\begin{figure}[htb]
\centering
\mbox{
{\scalebox{0.23}{\includegraphics{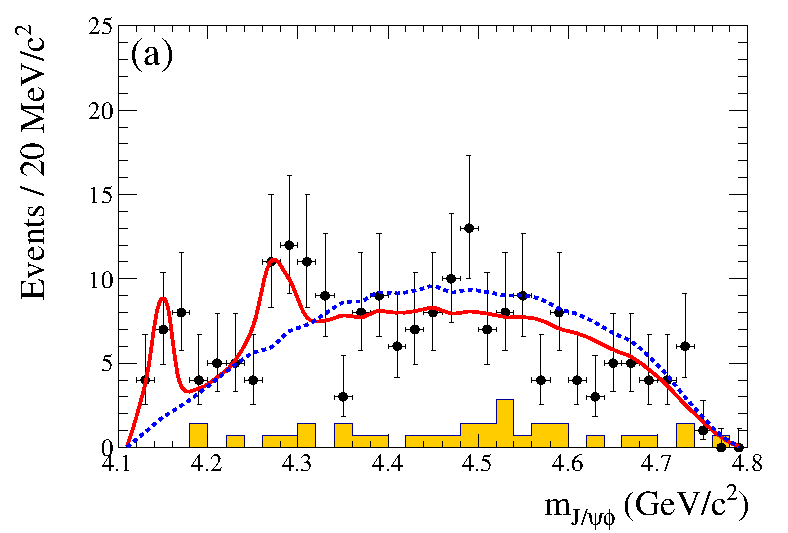}}} \quad
{\scalebox{0.23}{\includegraphics{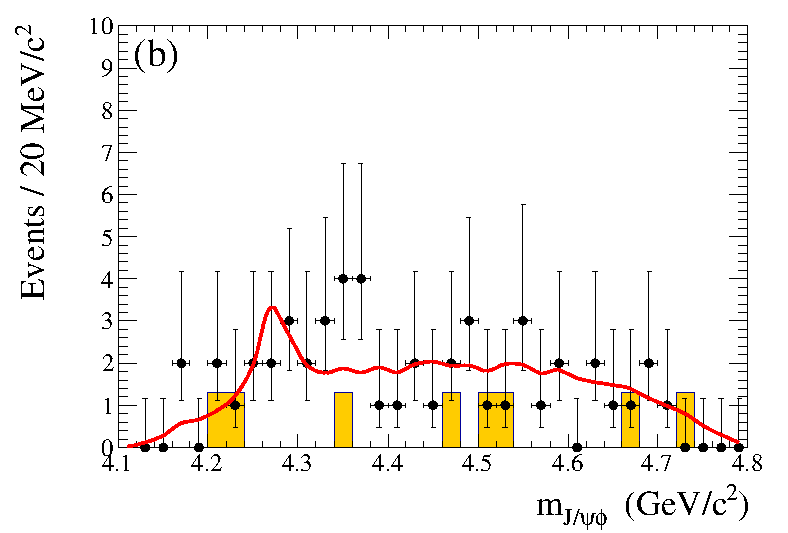}}} \quad
{\scalebox{0.23}{\includegraphics{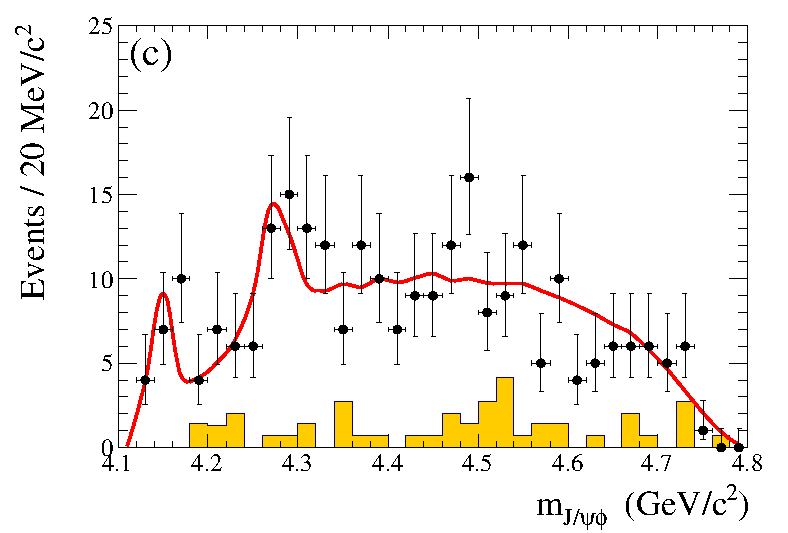}}}
}
\label{Fig3-babar}
\caption{Projections on the $J/\psi \phi$ mass spectrum from the Dalitz plot fit with the $X(4140)$ and the $X(4270)$ resonances for the (a) $B^+$, (b) $B^0$, and (c) combined $B^+$ and  $B^0$ data samples. The continuous (red) curves result from the fit; the dashed (blue) curve in (a)  indicates the projection for fit model D, with no resonances. The shaded (yellow) histograms show the estimated background contributions.}
\end{figure}

 We observed significant efficiency decrease at low $J/\psi \phi$ mass (see Fig.~1), due to the inability to reconstruct slow kaons in the laboratory frame, as a result of energy loss in the beampipe and SVT material. We perform an unbinned maximum likelihood fit to the channel $B^+ \rightarrow J/\psi \phi K^+$, modeling the resonances with an inchoerent sum of two S-wave relativistic Breit-Wigner (BW) functions with parameters fixed to the CDF values~\cite{kai}. A non-resonant contribution is described according to PHSP.  The decay of a pseudoscalar meson to two vector states contains high spin contributions which could generate non-uniform angular distributions. However, due to the limited data sample (212 yield for $B^+$ and 50 for $B^0$, in the signal area, respectively) we do not include such angular terms, and assume that the resonances decay isotropically. The fit function is  weighted by the inverse of the two-dimentional efficiency computed on the Dalitz plots (see the continuous red curve in Fig.~2-3).
 
We perform the fits using models with two resonances (Fig.~2-3), one resonance, and no resonances. All models provide a reasonably good description of the data, with $\chi^2$ probability larger than 5$\%$. We obtain the following corrected-estimates for the fractions for $B^+$, where the central values of mass and width of the two resonances are fixed to the values recently published from CDF\cite{kai} (Eq.~1) and CMS~\cite{cms} (Eq.~2), respectively:\\
\begin{eqnarray}
f_{X(4140)} = (9.2 \pm 3.3 \pm 4.7)\%,   f_{X(4270)} = (10.6 \pm 4.8 \pm 7.1) \%,  \end{eqnarray} 
\begin{eqnarray}
f_{X(4140)} = (13.2 \pm 3.8 \pm 6.8)\%,  f_{X(4270)} = (10.9 \pm 5.2 \pm 7.3) \%.
\end{eqnarray}

These values are consistent with each others within the uncertainties.
For comparison, CMS reported a fraction of $0.10 \pm 0.03$ for the X(4140), compatible with CDF, LHCb and our values within the uncertainties. CMS could not determine reliably the significance of the second structure X(4270) due to possible reflections of two-body decays. A significance smaller than 2$\sigma$ is found for the 2 peaks, within systematic uncertainties. Using the Feldman-Cousins method\cite{FC}, we obtain the ULs at 90\% CL:\\
\begin{eqnarray}
BF(B^+ \rightarrow X(4140)K^+)\times BF(X(4140) \rightarrow J/\psi \phi)/BF(B^+ \rightarrow J/\psi \phi K^+) < 0.135 ,
\end{eqnarray}
\begin{eqnarray}
BF(B^+ \rightarrow X(4270)K^+)\times BF(X(4270)\rightarrow J/\psi \phi)/BF(B^+ \rightarrow J/\psi \phi K^+) < 0.184.
\end{eqnarray}

The $X(4140)$ limit may be compared with the CDF measurement of $0.149\pm 0.039\pm 0.024$~\cite{kai} and the LHCb limit of 0.07~\cite{LHCb}. The $X(4270)$ limit may be compared with the LHCb limit of 0.08.

A detailed description of all BFs and ULs shortly introduced in this report is provided  in Ref.~\cite{elisabetta}: this work has been recently submitted to PRD.

As an additional contribution to Ref.~\cite{elisabetta}, which explains in detail this analysis performed by $BABAR$, we provide some plots for a comparison between the $BABAR$ data and the data published in Ref.~\cite{kai,cms, LHCb, d0}. In Fig.~4 you can see a comparison among the data of other experiments that published on the $J/\psi \phi$ invariant mass for $B^+ \rightarrow J/\psi \phi K^+$: the data are scaled by a factor taking in consideration the different integrated luminosity used from each experiment, and they are also background-subtracted, using information from Ref.~\cite{kai,cms, LHCb, d0}. We also rebin properly the data in the histograms of Fig.~4-5, for the correct comparison. Fig.~5 shows the comparison between the $BABAR$ data, re-weighted by the Dalitz efficiency and background-subtracted,  and what was published from other experiments.
 
The other experiments investigate only the $B^+$ decay within a limited $J/\psi \phi$ mass region, which is different for each experiment, while $BABAR$ analyzed the full range of $J/\psi \phi$ for both decay modes, $B^0$ and $B^+$, where $J/\psi \rightarrow e^+ e^-$ and  $J/\psi \rightarrow \mu^+ \mu^-$. For the other experiments the BFs are not estimated, so we cannot do comparison between our BF measurements and the others.

\begin{figure}[htb]
\centering
\mbox{
{\scalebox{0.25}{\includegraphics{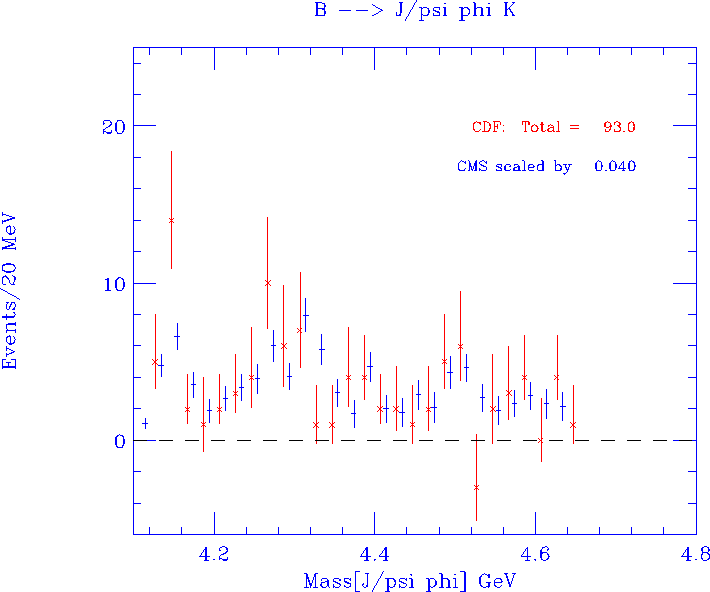}}} \quad
{\scalebox{0.25}{\includegraphics{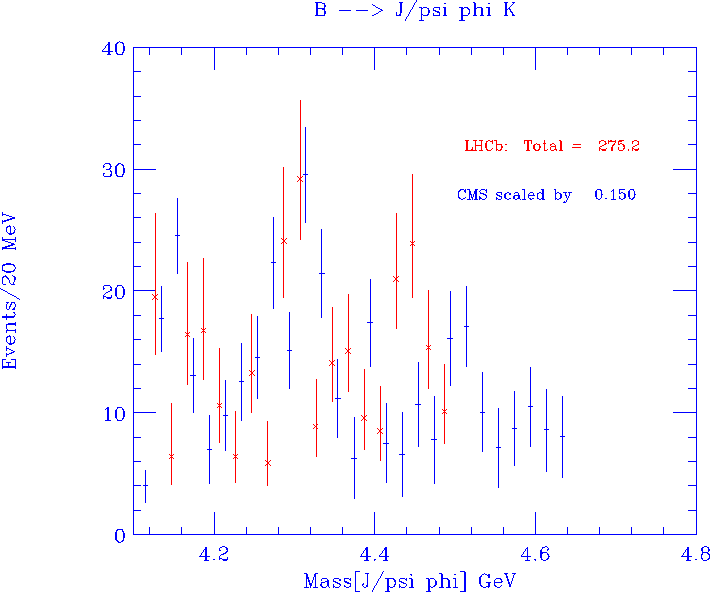}}}\quad
{\scalebox{0.25}{\includegraphics{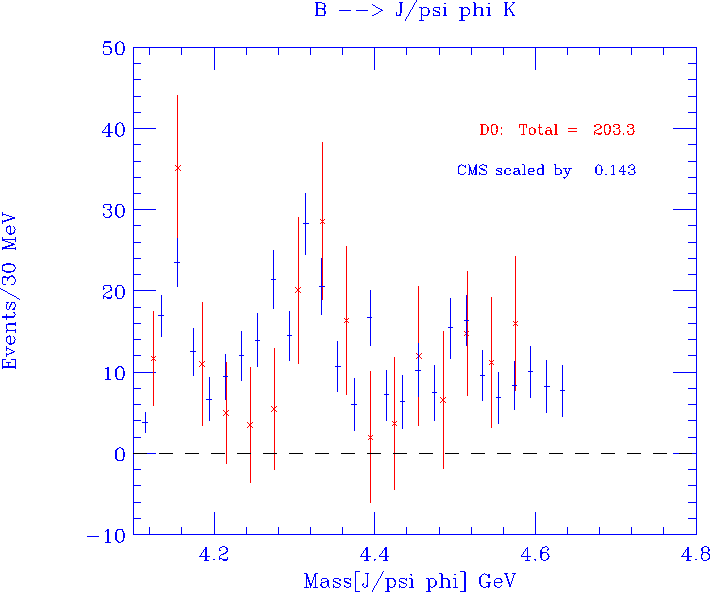}}}
}
\label{Figcomp8-babar}
\caption{$J/\psi \phi$ invariant mass distributions for the comparison between the CDF and CMS data (left); LHCb and CMS data (center); D0 and CMS data (right). Informations are taken from the public references ~\cite{kai, LHCb, cms, d0}. Data are rebinned, background-subtracted and properly scaled as indicated in the labels, for the comparison.}
\end{figure}

\begin{figure}[htb]
\centering
\mbox{
{\scalebox{0.25}{\includegraphics{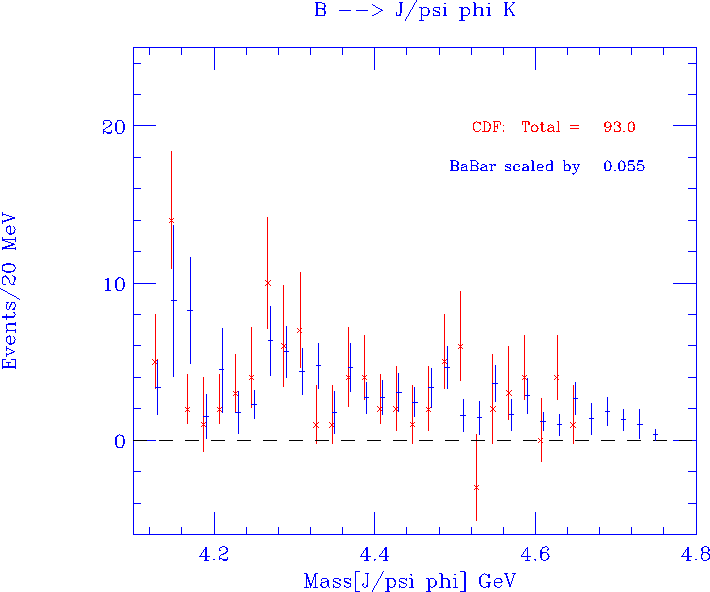}}} \quad
{\scalebox{0.25}{\includegraphics{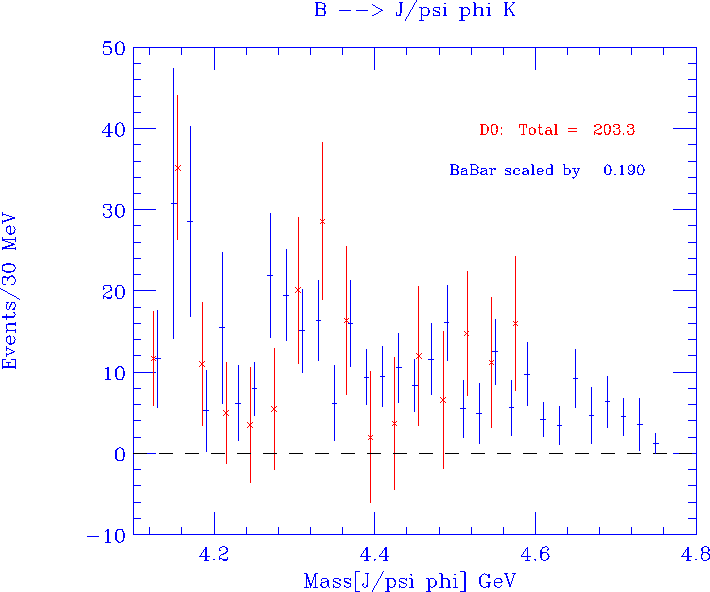}}}
}
\mbox{
{\scalebox{0.25}{\includegraphics{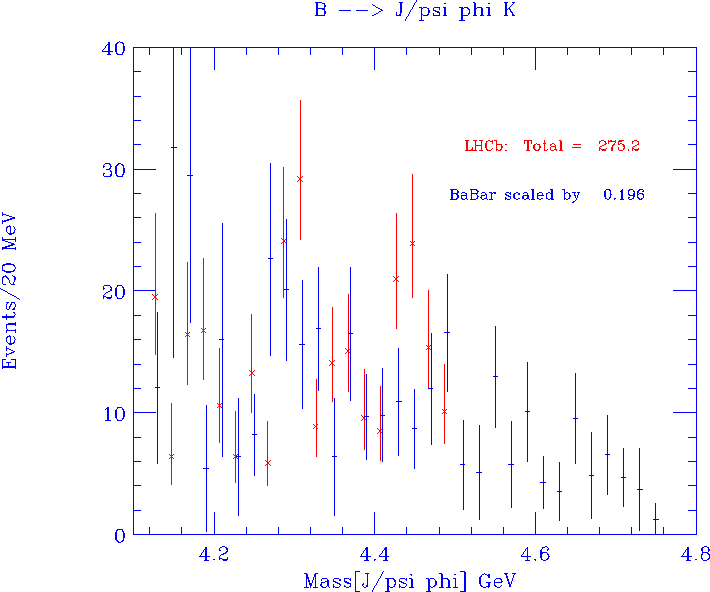}}}\quad
{\scalebox{0.25}{\includegraphics{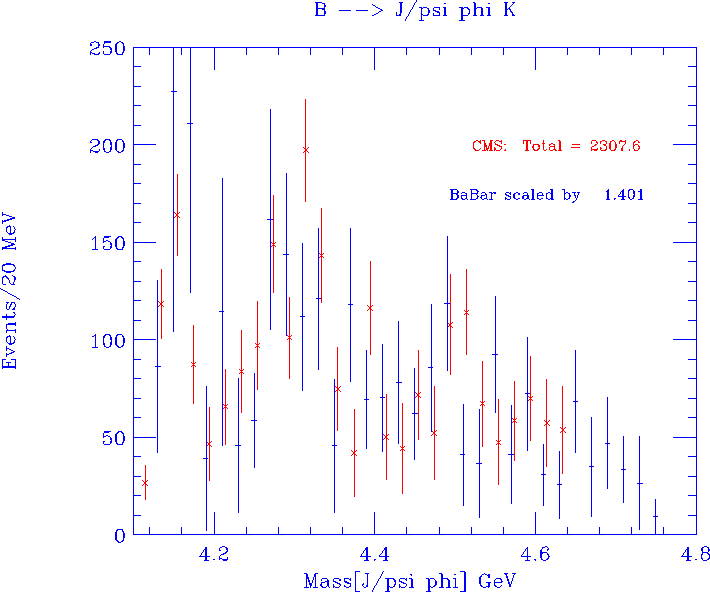}}}
}
\label{Figcomp7-babar}
\caption{Invariant mass system of $J/\psi \phi$: comparison between the $BABAR$ data (blue data points), reweighted by the Dalitz efficiency and background-subtracted,  and those published by CDF (top-left), D0 (top-right), LHCb (down-left), and CMS (down-right). Informations are taken from the references ~\cite{kai, LHCb, cms, elisabetta, d0}.}
\end{figure}

\section{Summary}
In summary, we observed signal for the decays $B^{+} \rightarrow J/\psi K^+ K^- K^{+}$,  $B^{0} \rightarrow J/\psi K^+ K^- K^0_S$, $B^{+} \rightarrow J/\psi \phi K^{+}$ and $B^{0} \rightarrow J/\psi \phi K^0_S$, obtaining currently the most precise BF measurements. We search for resonance production in the $J/\psi \phi$ mass spectrum and obtain significances below 2$\sigma$ for both the $X(4140)$ and the $X(4270)$ resonances, within systematic uncertainties.  Limits on the BF of these resonances are obtained. We find that the hypothesis that the events are distributed uniformly on the Dalitz plot gives a poorer description of the data. We also search for $B^0 \rightarrow J/\psi \phi$ and derive an UL on the BF for this decay mode, which is in agreement with theoretical expectations. The invariant mass distributions of $J/\psi K$ and $KKK$ do not show peaking structures.

\end{document}